\documentclass[aps,eqsecnum,preprint,floats,epsf,epsfig,nofootinbib,letter]{revtex4}
\usepackage{epsfig}
\usepackage{multirow}

\begin{document}
\def\be{\begin{eqnarray}}
\def\en{\end{eqnarray}}
\def\non{\nonumber}
\def\la{\langle}
\def\ra{\rangle}
\def\nc{N_c^{\rm eff}}
\def\vp{\varepsilon}
\def\drho{\bar\rho}
\def\deta{\bar\eta}
\def\CP{{\it CP}~}
\def\a{{\cal A}}
\def\B{{\cal B}}
\def\c{{\cal C}}
\def\d{{\cal D}}
\def\e{{\cal E}}
\def\p{{\cal P}}
\def\t{{\cal T}}
\def\up{\uparrow}
\def\dw{\downarrow}
\def\vma{{_{V-A}}}
\def\vpa{{_{V+A}}}
\def\smp{{_{S-P}}}
\def\spp{{_{S+P}}}
\def\J{{J/\psi}}
\def\ov{\overline}
\def\Lqcd{{\Lambda_{\rm QCD}}}
\def\pr{{Phys. Rev.}~}
\def\prl{{Phys. Rev. Lett.}~}
\def\pl{{Phys. Lett.}~}
\def\np{{Nucl. Phys.}~}
\def\zp{{Z. Phys.}~}
\def\lsim{ {\ \lower-1.2pt\vbox{\hbox{\rlap{$<$}\lower5pt\vbox{\hbox{$\sim$}
}}}\ } }
\def\gsim{ {\ \lower-1.2pt\vbox{\hbox{\rlap{$>$}\lower5pt\vbox{\hbox{$\sim$}
}}}\ } }

\font\el=cmbx10 scaled \magstep2{\obeylines\hfill September, 2017}

\vskip 1.5 cm

\centerline{\large\bf Masses of Scalar and Axial-Vector $B$ Mesons Revisited}

\bigskip
\centerline{\bf Hai-Yang Cheng$^{1}$, Fu-Sheng Yu$^{2}$}
\medskip
\centerline{$^1$ Institute of Physics, Academia Sinica}
\centerline{Taipei, Taiwan 115, Republic of China}
\medskip
\centerline{$^2$ School of Nuclear Science and Technology, Lanzhou University} \centerline{Lanzhou 730000, People's Republic of China}
\medskip

\bigskip
\bigskip
\bigskip
\bigskip
\bigskip
\centerline{\bf Abstract}
\bigskip
\small

The SU(3) quark model encounters a great challenge in describing even-parity mesons. Specifically, the $q\bar q$ quark model has difficulties in understanding the light scalar mesons below 1 GeV, scalar and axial-vector charmed mesons and $1^+$ charmonium-like state $X(3872)$.  A common wisdom for the resolution of these difficulties lies on the coupled channel effects which will distort the quark model calculations. In this work, we focus on the near mass degeneracy of scalar charmed mesons, $D_{s0}^*$ and $D_0^{*0}$, and its implications. Within the framework of heavy meson chiral perturbation theory, we show that near degeneracy can be qualitatively understood as a consequence of self-energy effects due to strong coupled channels. Quantitatively, the closeness of $D_{s0}^*$ and $D_0^{*0}$ masses can be implemented by adjusting two relevant strong couplings and the renormalization scale appearing in the loop diagram. Then this in turn implies the mass similarity of $B_{s0}^*$ and $B_0^{*0}$ mesons.
The $P_0^* P'_1$ interaction with the Goldstone boson is crucial for understanding the phenomenon of  near degeneracy.
Based on heavy quark symmetry in conjunction with corrections from QCD and $1/m_Q$ effects, we obtain the masses of $B^*_{(s)0}$ and $B'_{(s)1}$ mesons, for example, $M_{B_{s0}^*}= (5715\pm1)\,{\rm MeV}+\delta\Delta_S$, $M_{B'_{s1}}=(5763\pm1)\,{\rm MeV}+\delta\Delta_S$ with $\delta\Delta_S$ being $1/m_Q$ corrections.
We find that the predicted mass difference of 48 MeV between $B'_{s1}$ and $B_{s0}^*$ is larger than that of $20\sim 30$ MeV inferred from the relativistic quark models, whereas the difference of 15 MeV between the central values of $M_{B'_{s1}}$ and $M_{B'_1}$ is much smaller than the quark model expectation of $60-100$ MeV. Experimentally, it is important to have a precise mass measurement of $D_0^*$ mesons, especially the neutral one, to see if the non-strange scalar charmed meson is heavier than the strange partner as suggested by the recent LHCb measurement of the $D_0^{*\pm}$.

\pagebreak

\section{Introduction}
Although the SU(3) quark model has been applied successfully to describe the properties of hadrons such as pseudoscalar and vector mesons, octet and decuplet baryons, it often encounters a great challenge in understanding even-parity mesons, especially scalar ones.  Take vector mesons as an example and consider the octet vector ones: $\rho, \omega, K^*,\phi$. Since the constituent strange quark is heavier than up or down quark by 150 MeV, one will expect the mass hierarchy pattern $m_\phi> m_{K^*}>m_\rho\sim m_\omega$ which is borne out by experiment. However, this quark model picture faces great challenges in describing the even-parity meson sector:

\begin{itemize}

\item
Many scalar mesons with masses lower than 2 GeV have been observed and they can be classified into two nonets: one nonet with mass below or close to 1 GeV, such as $f_0(500)$ (or $\sigma$), $K_0^*(800)$ (or $\kappa$), $f_0(980)$ and $a_0(980)$  and the other nonet with mass above 1 GeV such as $K_0^*(1430)$, $a_0(1450)$ and two isosinglet scalar mesons. Of course, the two nonets cannot be both low-lying $^3P_0$ $q\bar q$ states simultaneously. If the light scalar nonet is identified with the P-wave $q\bar q$ states, one will encounter two major difficulties: First, why are $a_0(980)$ and $f_0(980)$ degenerate in their masses? In the $q\bar q$ model, the latter is dominated by the $s\bar s$ component, whereas the former cannot have the $s\bar s$ content since it is an $I=1$ state. One will expect the mass hierarchy pattern : $m_{f_0(980)}>m_{K_0^*(800)}>m_{a_0(980)}\sim m_{f_0(500)}$. However, this pattern is not seen by experiment. In contrast, it is $m_{a_0(980)}\approx m_{f_0(980)}>m_{K_0^*(800)}> m_{f_0(500)}$ experimentally. Second, why are $f_0(500)$ and $K_0^*(800)$ so broad compared to the narrow widths of $a_0(980)$ and $f_0(980)$ even though they are all in the same nonet?

\item
In the scalar meson sector above 1 GeV, $K_0^*(1430)$ with mass $1425\pm 50$ MeV \cite{PDG}  is almost degenerate in masses with $a_0(1450)$ which has a mass of $1474\pm19$ MeV \cite{PDG} despite having one strange quark for the former.

\item
In the even-parity charmed meson sector, we compare the experimentally measured masses and widths with what are expected from the quark model (see Table \ref{tab:pwavecharm}). There are some prominent features from this comparison: (i) The measured masses of $D_0^*, D'_1,D_{s0}^*$ and $D'_{s1}$ are substantially smaller than the quark model predictions. (ii) The physical $D_{s0}^*$ mass is below the $DK$ threshold, while $D'_{s1}$ is below $DK^*$. This means that both of them are quite narrow, in sharp contrast to the quark model expectation of large widths for them. (iii) $D_0^*(2400)^0$ and $D_{s0}^*(2317)$ are almost equal in their masses, while $D_0^*(2400)^\pm$ is heavier than $D_{s0}^*$ even though the latter contains a strange quark.
\footnote{Strictly speaking, the masses of the neutral and charged states of $D_0^*(2400)$ are not consistently determined due mainly to its broadness. The mass of the charged one is primarily from the LHCb measurements \cite{LHCb:D0pm}, while the neutral one is from  BaBar \cite{BaBar:D0}, Belle \cite{Belle:D0}  and FOCUS \cite{FOCUS:D0}. It is worthwhile to notice that only FOCUS has measured both the neutral and charged $D_0^*(2400)$, and their masses are quite similar with a small difference of a few MeV. All the other three groups have not reported the masses for both the neutral and charged $D_0^*(2400)$. In addition, the masses reported by different groups are very different from each other. As a result, the world averaged masses for the neutral and charged $D_0^*(2400)$ shown in Table \ref{tab:pwavecharm} are very different as well. The difference is well beyond the expectation from isospin splitting. Thus, the two averaged masses given by the Particle Data Group (PDG) \cite{PDG} are not consistent with each other. In this work, we consider both PDG masses for $D_0^*(2400)$. In Sec. II we focus on the closeness of $D_{s0}^*$ and $D_0^*$ masses and discuss its implication to the scalar $B$ sector. In Sec. III we derive two different sets of scalar $B$ meson masses, corresponding to two different PDG masses for $D_0^*(2400)$.}
(iv) The masses of $D_1,D_2^*, D_{s1}$ and $D_{s2}^*$ predicted by the quark model are consistent with experiment. These four observations lead to the conclusion that $0^+$ and $1^{\prime +}$ charmed mesons have very unusual behavior not anticipated from the quark model.

\item
The first $X\!Y\!Z$ particle, namely $X(3872)$,  observed by Belle in 2003 in $B^\pm\to K^\pm + (J/\psiĎ\pi^+\pi^-)$ decays \cite{Belle:X}, has the quantum numbers $J^{PC}=1^{++}$ \cite{LHCb:X}.
$X(3872)$ cannot be a pure charmonium as it cannot be identified as $\chi_{c1}(1^3P_1)$ with a mass 3511 MeV \cite{PDG} or $\chi_{c1}(2\,^3P_1)$ with the predicted mass of order 3950 MeV \cite{Godfrey:1985xj}. Moreover, a pure charmonium for $X(3872)$ cannot explain the large isospin violation observed in $X(3872)\to J/\psi\omega, J/\psi\rho$ decays.
The extreme proximity of $X(3872)$ to the threshold suggests a loosely
bound molecule state $D^0\bar D^{*0}$ for $X(3872)$. On the other hand, $X(3872)$ cannot be a pure $D\!\bar D^*$  molecular state either for the following reasons: (i) It cannot explain the prompt production of $X(3872)$  in high energy collisions \cite{Suzuki:2005ha,Braaten:2009wk}. (ii) The ratio $R_1 \equiv \Gamma(B^0\to K^0X(3872))/ƒ·\Gamma(B^-\to K^-X(3872))$ is predicted to be much less than unity in the molecular scenario, while it was measured to be $0.50\pm0.30\pm0.05$ and $1.26\pm0.65\pm0.06$  by Belle \cite{Belle:Xcs}.
(iii) For the ratio $R_2\equiv\Gamma(X(3872)\to \psi(2S)\gamma)/\Gamma(X(3872)\to J/\psi(1S)\gamma)$,  the molecular model leads to a very small value of order $3\times 10^{-3}$ \cite{Swanson:2006st,Dong:2009uf,Ferretti:2014xqa} while the charmonium model predicts $R_2$ to be order of unity. The  LHCb measurement yields $R_2 = 2.46\pm0.64\pm0.29$ \cite{LHCb:Xgamma}. Hence, $X(3872)$ cannot be a pure $D\!\bar D^*$ molecular state.

The above discussions suggest that $X(3872)$ is most likely an admixture of the $S$-wave $D\!\bar D^*$ molecule and the $P$-wave charmonium as first advocated in \cite{Suzuki:2005ha}
\be \label{eq:X}
|X(3872)\ra &=& c_1|c\bar c\ra_{\rm P-wave}+ c_2|D^0\ov D^{*0}\ra_{\rm S-wave} + c_3|D^+ D^{*-}\ra_{\rm S-wave}+\cdots.
\en
More specifically, the charmonium is identified with  $\chi_{c1}(2\,^3P_1)$. Some calculations favor a larger $c\bar c$ component over the $D^0\ov D^{*0}$ component (see e.g. \cite{Matheus,Kang}). Then the question is how to explain the mass of $X(3872)$ through the charmonium picture.

\end{itemize}

\begin{table}[t] \label{tab:pwavecharm}
\centering \caption{Measured masses and widths of even-parity charmed mesons. The four $p$-wave charmed meson states are denoted by
$D_0^*,D'_1,D_1$ and $D_2^*$, respectively. In the heavy quark limit, $D'_1$ has
$j=1/2$ and $D_1$ has $j=3/2$ with $j$ being the total angular
momentum of the light degrees of freedom. The data are taken from the Particle Data Group \cite{PDG}. The last two columns are the predictions from the quark model \cite{Godfrey,Di Pierro}. ``Large" means a broad width of order 100 MeV, while ``small" implies a narrow width of order 10 MeV. }
\vskip 0.5cm
\begin{tabular}{|c c  c c |c c| }\hline
~~$J^P$ & Meson  &   Mass (MeV)  & $\Gamma$ (MeV)  & ~~~Mass (MeV) & ~~$\Gamma$~~~~~~  \\ \hline
$0^+$ & $D_0^{*}(2400)^0$ & $2318 \pm 29$ & $267 \pm 40$ & 2340$-$2410 & large \\
& $D_0^{*}(2400)^\pm$ & $ 2351 \pm 7 $ & $~230 \pm 17~$ & 2340$-$2410 & large \\
$1^{\prime+}$ & $D_{1}^{\prime  }(2430)^0$ & $ 2427 \pm 26\pm 25$ & $384^{+107}_{-~75}  \pm 74$ & 2470$-$2530 & large \\
$1^+$ & $D_1(2420)^0$ & $2420.8\pm0.5$ & $31.7\pm2.5$ & 2417$-$2434 & small \\
& $D_1(2420)^\pm$ & $2432.2\pm2.4$ & $25\pm6$ & 2417$-$2434 & small \\
$2^+$ & $D_2^*(2460)^0$ & $2460.57\pm0.15$ & $47.7\pm1.3$ & 2460$-$2467 & small \\
 & $D_2^*(2460)^\pm$ & $2465.4\pm1.3$ & $46.7\pm1.2$ & 2460$-$2467 & small \\ \hline
$0^+$ & $D_{s0}^{*}(2317)^\pm$ & $2317.7 \pm 0.6 $ & $<3.8$ &2400$-$2510  & large  \\
$1^{\prime+}$ & $D_{s1}^{\prime}(2460)^\pm$ & $2459.5 \pm 0.6$ & $<3.5$ & 2528$-$2536 & large \\
$1^+$ & $D_{s1}(2536)^\pm$ & $2535.10\pm0.06$ & $0.92\pm0.05$ & 2543$-$2605 & small  \\
$2^+$ & $D_{s2}^*(2573)^\pm$ & $2569.1\pm0.8$ & $16.9\pm0.8$ & 2569$-$2581 & small \\ \hline
\end{tabular}
\end{table}

In short, the $q\bar q$ quark model has difficulties in describing light scalar mesons below 1 GeV, $0^+$ and $1'^{+}$ charmed mesons and $1^+$ charmonium-like state $X(3872)$.  A common wisdom for the resolution of aforementioned difficulties lies on the coupled channel effects which will distort the quark model calculations.

In the quark potential model, the predicted masses for $D_{s0}^*$ and $D_0^{*0}$  are higher than the measured ones by order 160 MeV and 70 MeV, respectively \cite{Godfrey,Di Pierro}.
It was first stressed and proposed in \cite{vanBeveren} that the low mass of $D_{s0}^*(2317)$ ($D_0^*(2400)^0$) arises from the mixing between the $0^+$ $c\bar s$ ($c\bar q$) state and the $DK$ ($D\pi$) threshold  (see also \cite{Browder}). This conjecture was realized in both QCD sum rule \cite{Dai,Dai:D0} and lattice \cite{Mohler:Ds0,Lang,Mohler:D0} calculations.
For example, when the contribution from the $DK$ continuum is included in QCD sum rules, it has been shown that this effect will significantly lower the mass of the $D_{s0}^*$ state \cite{Dai}. Recent lattice calculations using $c\bar s$, $DK$ and $D^*K$ interpolating fields show the existence of $D_{s0}^*(2317)$ below the $DK$ threshold \cite{Mohler:Ds0} and $D'_{s1}(2460)$ below the $D^*K$ threshold \cite{Lang}.
\footnote{A recent lattice calculation with $N_f=2+1+1$ optimal domain-wall fermions \cite{Chen:2017kxr} yields a mass of $2317\pm15\pm5$ MeV for $D_{s0}^*$ and $2463\pm13\pm9$ MeV for $D'_{s1}(2460)$.
}
All these results indicate that the strong coupling of scalars with hadronic channels will play an essential role of lowering their masses.

By the same token, mass shifts of charmed and bottom scalar mesons due to self-energy hadronic loops have been calculated in \cite{Guo}. The results imply that the bare masses of scalar mesons calculated in the quark model can be reduced significantly. Mass shifts due to hadronic loops or strong coupled channels have also been studied in different frameworks to explain the small mass of $D_{s0}^*(2317)$ \cite{Hwang:2004cd,Zhou,TLee,Ortega:2016mms}.
In the same spirit, even if $X(3872)$ is dominated by the $c\bar c$ component, the mass of $\chi_{c1}'$ can be shifted down due to its strong coupling with $D\!\bar D^*$ channels \cite{Li:2009ad,Danilkin:2010cc,Achasov:2015oia}.

Both $f_0(980)$ and $a_0(980)$ have the strong couple channel $K\ov K$. They are
often viewed as $K\ov K$  molecules, which accounts for their
near degeneracy with $2m_K$. Schematically, the self-energy $K\ov K$ loop diagram of $a_0(980)$ will shift its mass to the physical one. In the unitarized chiral perturbation theory, light scalar mesons $f_0(500)$, $K_0^*(800)$, $f_0(980)$ and $a_0(980)$ can be dynamically generated through their strong couplings with $\pi\pi$, $K\pi$, $K\ov K$, and $K\ov K$, respectively \cite{lightS}.
Alternatively, it is well known that the tetraquark picture originally advocated by Jaffe \cite{Jaffe} provides a simple solution to the mass and width hierarchy problems in the light scalar meson sector. The tetraquark structure of light scalars accounts for the mass hierarchy pattern $m_{a_0(980)}\approx m_{f_0(980)}>m_{K_0^*(800)}> m_{f_0(500)}$, Moreover, the $S$-wave 4-quark nonet can be lighter than the $P$-wave $q\bar q$ nonet above 1 GeV due to the absence of the orbital angular momentum barrier and the presence of strong attraction between the diquarks $(qq)_{\bf 3^*}$ and $(\bar q\bar q)_{\bf 3}$ \cite{Jaffe2}. The fall-apart decays $f_0(500)\to \pi\pi$, $K_0^*(800)\to \pi K$, and $f_0(980), a_0(980)\to K\ov K$
are all OZI-superallowed. This
explains the very broad widths of $f_0(500)$ and $K_0^*(800)$, and the narrowness of $f_0(980)$ and $a_0(980)$ owing to the very limited phase space available as they are near the $K\ov K$ threshold.

In \cite{Cheng:2014} we have studied near mass degeneracy of scalar charmed
and bottom mesons.
Qualitatively, the approximate mass degeneracy can be understood as a consequence of self-energy effects due to strong coupled channels which will push down the mass of the heavy scalar meson in the strange sector more than that in the non-strange partner. However,  we showed that  it works in the conventional model without heavy quark expansion, but not in the approach of heavy meson chiral perturbation theory (HMChPT) as mass degeneracy and the physical masses of $D_{s0}^*$ and $D_0^*$ cannot be accounted for simultaneously. Mass shifts in the strange charm sector are found to be largely overestimated. It turns out that the conventional model works better toward the understanding of near mass degeneracy.

Our previous work was criticized by Alhakami \cite{Alhakami} who followed the framework of \cite{Mehen} to write down the general expression of HMChPT  and fit the unknown low-energy constants in the effective Lagrangian to the experimentally measured odd- and even-parity charmed mesons. Using the results from the charm sector, Alhakami predicted the spectrum of odd- and even-parity bottom mesons. He concluded that the near degeneracy of nonstrange and strange scalar $B$ mesons is confirmed in the predictions using HMChPT. He then proceeded to criticize that we should use physical masses instead of bare masses to evaluate the hardronic loop  effects and that we have missed the contributions from axial-vector heavy mesons to the self-energy of scalar mesons.

Motivated by the above-mentioned criticisms \cite{Alhakami}, in this work we shall re-examine our calculations within the framework of HMChPT. We show that the closeness of $D_{s0}^*$ and $D_0^{*0}$ masses can be achieved by taking into account the additional contribution, which was missing in our previous work, from axial-vector heavy mesons to the self-energy diagrams of scalar mesons
by adjusting two relevant strong couplings and the renormalization scale $\mu$ appearing in the loop diagram.  Then we proceed to confirm that near degeneracy observed in the charm sector will imply the similarity of $B_{s0}^*$ and $B_0^*$ masses in the $B$ system.

This work is organized as follows.  In Sec. II we consider the self-energy corrections to scalar and axial-vector heavy mesons in HMChPT. In the literature, the self-energy loop diagrams were sometimes evaluated in HMChPT by neglecting the corrections from mass splittings and residual masses to the heavy meson's propagator. We shall demonstrate in Sec. III that
the calculation in this manner does not lead to the desired degeneracy in both charm and $B$ sectors simultaneously. The masses of $B_{s0}^*$ and $B_{0}^*$ are discussed in Sec. IV with focus on the predictions based on heavy quark symmetry and possible $1/m_Q$ and QCD corrections. Sec. V comes to our conclusions.

\section{Mass shift of scalar and axial-vector heavy mesons due to hadronic loops}

Self-energy hadronic loop corrections to $0^+$ and $1^{\prime+}$ heavy mesons have been considered in the literature \cite{Becirevic,Fajfer,Brdnik,Mehen,Alhakami,Ananth,Guo,Cheng:2014}.
Since chiral loop corrections to heavy scalar mesons have both finite and divergent parts, it is natural to consider the framework of HMChPT where the divergences and the renormalization scale dependence arising from the chiral loops induced by the lowest-order tree Lagrangian can be absorbed into the counterterms which have the same structure as the next-order tree Lagrangian.

The heavy meson's propagator in HMChPT has the expression
\be
{i\over 2v\cdot k-\Pi(v\cdot k)},
\en
where $v$ and $k$, respectively, are the velocity and  the residual momentum of the meson defined by $p=vm_0+k$, and $\Pi(v\cdot k)$ is the 1PI self-energy contribution. In general,  $\Pi(v\cdot k)$ is complex as its imaginary part is related to the resonance's width.
The particle's on-shell condition is then given by
\be \label{eq:onshell}
2v\cdot \tilde k-{\rm Re}\Pi(v\cdot \tilde k)=0.
\en
The physical mass reads
\be \label{eq:massPi}
m=m_0+v\cdot \tilde k=m_0+{1\over 2} \,{\rm Re}\Pi(v\cdot \tilde k)\,.
\en

\begin{figure}[t]
\begin{center}
\includegraphics[width=0.70\textwidth]{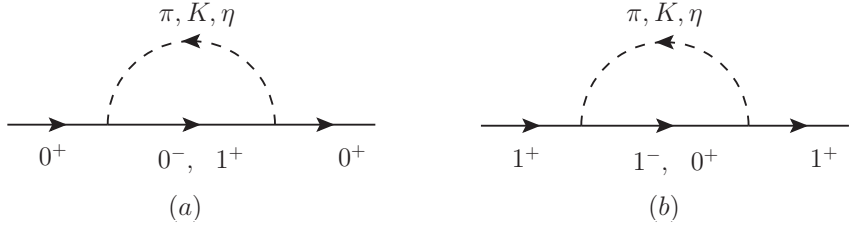}
\vspace{0.0cm}
\caption{Self-energy contributions to scalar and axial-vector heavy mesons.} \label{fig:Scalar} \end{center}
\end{figure}

Consider the self-energy diagrams depicted in Fig. \ref{fig:Scalar} for scalar and axial-vector heavy mesons. We will evaluate the loop diagrams in the framework of HMChPT in which the low energy dynamics of hadrons is described by the formalism in which heavy quark symmetry and chiral symmetry are synthesized \cite{Yan,Wise,Donoghue}. The relevant Lagrangian is \cite{Casalbuoni}
\be \label{eq:Lagrangian}
{\cal L} &=& {\rm Tr}\left[ \bar H_b(iv\cdot D)_{ba}H_a\right]+
{\rm Tr}\left[\bar S_b((iv\cdot D)_{ba}-\delta_{ba}\Delta_S)S_a\right] \non \\
&+& g{\rm Tr}\left[\bar H_b\gamma_\mu\gamma_5 {\cal A}^\mu_{ba}H_a\right]
+h{\rm Tr}\left[\bar S_b\gamma_\mu\gamma_5 {\cal A}^\mu_{ba}H_a +h.c. \right]+g'{\rm Tr}\left[\bar S_b\gamma_\mu\gamma_5 {\cal A}^\mu_{ba}S_a +h.c. \right],
\en
where $H$ denotes the odd-parity spin doublet $(P,P^*)$ and $S$ the even-parity spin doublet $(P_0^*,P'_1)$ with $j=1/2$ ($j$ being the total angular momentum of the light degrees of freedom):
\be \label{eq:field}
H_a={ 1+v\!\!\!/\over 2}[P^*_{a\mu}\gamma^\mu-P_a\gamma_5], \qquad
S_a={1\over 2}(1+v\!\!\!/)[{P'}_{1a}^{\mu}\gamma_\mu\gamma_5-P^*_{0a}],
\en
with $a=u,d,s$, for example, $P_a=(D^0,D^+,D_s^+)$.
The nonlinear chiral symmetry is realized by making use of the unitary matrix $\Sigma={\rm exp}(i\sqrt{2}\phi/f_\pi)$ with $f_\pi=93$ MeV and $\phi$ being a $3\times 3$ matrix for the octet of Goldstone bosons.
In terms of the new matrix $\xi=\Sigma^{1/2}$, the axial vector field ${\cal A}$ reads ${\cal A}={i\over 2}(\xi^\dagger\partial_\mu\xi-\xi\partial_\mu \xi^\dagger)$.
In Eq. (\ref{eq:Lagrangian}), the parameter $\Delta_S$ is the residual mass of the $S$ field; it measures the mass splitting between even-
and odd-parity doublets and can be expressed in terms of the spin-averaged masses
\be
\la M_H\ra\equiv {3M_{P^*}+M_{P}\over 4},\qquad \la M_S\ra\equiv {3M_{P'_1}+M_{P^*_0}\over 4},
\en
so that
\be \label{eq:DeltaS}
\Delta_{S}= \la M_S\ra-\la M_H\ra.
\en

There exist two corrections to the chiral Lagrangian (\ref{eq:Lagrangian}): one from $1/m_Q$ corrections and the other from chiral symmetry breaking.
The $1/m_Q$ corrections are given by \cite{Casalbuoni,Boyd}
\be
{\cal L}_{1/m_Q}={1\over 2 m_Q}\left\{ \lambda_2^H{\rm Tr}[\bar H_a\sigma^{\mu\nu}H_a\sigma_{\mu\nu}]-\lambda_2^S {\rm Tr}[\bar S_a\sigma^{\mu\nu}S_a\sigma_{\mu\nu}]\right\},
\en
with
\be \label{eq:lambda2}
\lambda_2^H={1\over 4}(M^2_{P^*}-M^2_{P}), \qquad \lambda_2^S={1\over 4}(M^2_{P'_1}-M^2_{P^*_0}),
\en
where $\lambda_H$ ($\lambda_S$) is the mass splitting between spin partners, namely, $P^*$ and $P$ ($P'_1$ and $P^*_0$) of the pseudoscalar (scalar) doublet. We will not write down the explicit expressions for chiral symmetry breaking terms and the interested reader is referred to \cite{Mehen}.
The masses of heavy mesons can be expressed as
\be \label{eq:massrel}
M_{P_a} &=& M_0-{3\over 2}{\lambda_2^H\over m_Q}+\Delta_a, \qquad\qquad M_{P^*_a} = M_0+{1\over 2}{\lambda_2^H\over m_Q}+\Delta_a, \non \\
M_{P_{0a}^*} &=& M_0+\Delta_S-{3\over 2}{\lambda_2^S\over m_Q}+\tilde\Delta_a, \quad~ M_{P'_{1a}} = M_0+\Delta_S+{1\over 2}{\lambda_2^S\over m_Q}+\tilde\Delta_a,
\en
where $\Delta_a$ and $\tilde\Delta_a$ denote the residual mass contributions to odd- and even-parity mesons, respectively.
Note that $\lambda_2^H/m_Q\approx {1\over 2}(M_{P^*}-M_P)\equiv {1\over 2}\Delta M_P$ and $\lambda_2^S/m_Q\approx {1\over 2}(M_{P'_1}-M_{P_0^*})\equiv {1\over 2}\Delta M_S$ in the heavy quark limit. The propagators for $P_a(v)$, $P_a^*(v)$, $P_{0a}^*(v)$ and $P'_{1a}(v)$ read
\be \label{eq:Pprop}
{i\over 2(v\cdot k+{3\over 4}\Delta M_P-\Delta_a)+i\epsilon}, \qquad {-i(g_{\mu\nu}-v_\mu v_\nu)\over 2(v\cdot k-{1\over 4}\Delta M_P-\Delta_{a})+i\epsilon},
\en
and
\be \label{eq:Sprop}
{i\over 2(v\cdot k-\Delta_S +{3\over 4}\Delta M_S-\tilde\Delta_{a})+i\epsilon}, \qquad
{-i(g_{\mu\nu}-v_\mu v_\nu)\over 2(v\cdot k-\Delta_S -{1\over 4}\Delta M_S-\tilde\Delta_{a})+i\epsilon},
\en
respectively.

Consider the hadronic loop contribution to $D_{s0}^*$ in Fig. \ref{fig:Scalar}(a) with the intermediate states $D^0$ and $K^+$.
The self-energy loop integral is
\be \label{eq:loop}
\Pi_{DK}(\omega_D) &=& \left({2h^2\over f_\pi^2}\right){i\over 2}\int {d^4q\over (2\pi)^4}\,{(v\cdot q)^2\over (q^2-m_K^2+i\epsilon)(v\cdot k'+{3\over 4}\Delta M_D-\Delta_u+ i\epsilon)} \non \\
&=& \left({2h^2\over f_\pi^2}\right){i\over 2}\int {d^4q\over (2\pi)^4}\,{(v\cdot q)^2\over (q^2-m_K^2+i\epsilon)(v\cdot q+\omega_D+i\epsilon)},
\en
where $m$ is the mass of the Goldstone boson.
The residual momentum $k'$ of the heavy meson in the loop is given by
$k'=p+q-vM_D=q+k+v({\cal M}_{D_{s0}^*}-M_D)$,
and $\omega_D=v\cdot k+{\cal M}_{D_{s0}^*}-M_D+{3\over 4}\Delta M_D-\Delta_u$. The calligraphic symbol has been used to denote the bare mass.
The full $D_{s0}^*$ propagator becomes
\be \label{eq:Ds0prop}
{i\over 2(v\cdot k-\Delta_{S}+{3\over 4}\Delta M_{S}-\tilde\Delta_s) -[2\,\Pi_{DK}(\omega_D)+{2\over 3}\,\Pi_{D_s\eta}(\omega_{D_s})+2\,\Pi'_{D'_1K}(\omega'_{D'_1})+{2\over 3}\,\Pi'_{D'_{s1}\eta}(\omega'_{D'_{s1}})]} \non \\
\en
after taking into account the contributions from the channels $D^0K^+,D^+K^0,D_s^+\eta$  and $D_1^{\prime 0}K^+,D_1^{\prime +}K^0,D^{\prime+}_{s1}\eta$. In Eq. (\ref{eq:Ds0prop}),
\be \label{eq:loop2}
\Pi'_{D'_1K}(\omega'_{D'_1})
&=& -\left({2{g'}^2\over f_\pi^2}\right){i\over 2}\int {d^4q\over (2\pi)^4}\,{q^2-(v\cdot q)^2\over (q^2-m^2+i\epsilon)(v\cdot k'-\Delta_S-{1\over 4}\Delta M_S-\tilde \Delta_u+i\epsilon)} \non \\
&=& -\left({2{g'}^2\over f_\pi^2}\right){i\over 2}\int {d^4q\over (2\pi)^4}\,{q^2-(v\cdot q)^2\over (q^2-m^2+i\epsilon)(v\cdot q+\omega'_{D'_1}+i\epsilon)},
\en
where $\omega'_{D'_1}=v\cdot k+{\cal M}_{D_{s0}^*}-M_{D'_1}-(M_{D'_1}-M_D)+{3\over 4}\Delta M_D-\Delta_u$, and use of Eq. (\ref{eq:massrel}) has been made. Likewise, the full $D_{0}^*$ propagator reads
\be \label{eq:D0prop}
&&i\Bigg[2(v\cdot k-\Delta_{S}+{3\over 4}\Delta M_{S}-\tilde\Delta_u) -\Big[{3\over 2}\,\Pi_{D\pi}(\omega_D)+{1\over 6}\,\Pi_{D\eta}(\omega_{D})+\Pi_{D_sK}(\omega_{D_s}) \non \\
&& \qquad +{3\over 2}\,\Pi'_{D'_1\pi}(\omega'_{D'_1})+{1\over 6}\,\Pi'_{D'_{1}\eta}(\omega'_{D'_{1}})+\Pi'_{D'_{s1}K}(\omega'_{D'_{s1}})\Big]\Bigg]^{-1}.
\en

Since many parameters such as $\Delta_S$, $\Delta M_S$, $\tilde\Delta_s$ and $\tilde\Delta_u$ in Eqs. (\ref{eq:Ds0prop}) and (\ref{eq:D0prop}) are unknown, we are not able to determine mass shifts from above equations. Assuming that the bare mass ${\cal M}$ is the one obtained in the quark model, then from Eq. (\ref{eq:massrel}) we have
\be
\Delta_{S}-{3\over 4}\Delta M_S+\tilde\Delta_s={\cal M}_{D_{s0}^*}-M_0={\cal M}_{D_{s0}^*}-M_D-{3\over 4}\Delta M_D+\Delta_u.
\en
With
\be \label{eq:FforDs0}
F(v\cdot k)_{D_{s0}^*} &\equiv& 2(v\cdot k-{\cal M}_{D_{s0}^*}+M_D+{3\over 4}\Delta M_D-\Delta_u)-{\rm Re}\Big[2\,\Pi_{DK}(\omega_D)+{2\over 3}\,\Pi_{D_s\eta}(\omega_{D_s} ) \non \\
&& +2\,\Pi'_{D'_1K}(\omega'_{D'_1})+{2\over 3}\,\Pi'_{D'_{s1}\eta}(\omega'_{D'_{s1}})\Big], \non \\
F(v\cdot k)_{D_{0}^*} &\equiv& 2(v\cdot k-{\cal M}_{D_{0}^*}+M_D+{3\over 4}\Delta M_D-\Delta_u)-{\rm Re}\Big[{3\over 2}\,\Pi_{D\pi}(\omega_D)+{1\over 6}\,\Pi_{D\eta}(\omega_{D})+\Pi_{D_sK}(\omega_{D_s} ) \non \\
&& +{3\over 2}\,\Pi'_{D'_1\pi}(\omega'_{D'_1})+{1\over 6}\,\Pi'_{D'_{1}\eta}(\omega'_{D'_{1}})+\Pi'_{D'_{s1}K}(\omega'_{D'_{s1}})\Big],
\en
the on-shell conditions read $F(v\cdot \tilde k)_{D_{s0}^*}=0$ for $D_{s0}^*$ and $F(v\cdot \tilde k)_{D_{0}^*}=0$ for $D_{0}^*$.
The physical masses are then given by
\be \label{eq:physmass}
M_{D_{(s)0}^*}=M_0+v\cdot \tilde k=M_D+{3\over 4}\Delta M_D-\Delta_u+v\cdot \tilde k\,.
\en
Since $\Delta_u$ is of order 1 MeV \cite{Stewart}, it can be neglected in practical calculations. Note that in the above equation, one should not replace $M_0$ by the bare mass ${\cal M}_{D_{(s)0}^*}$. Indeed, in
the absence of chiral loop corrections,
$M_{D_{s0}^*}=M_0+v\cdot \tilde k=M_0+\Delta_S -{3\over 4}\Delta M_S+\tilde\Delta_s={\cal M}_{D_{s0}^*}$, as it should be.

For the self-energy of the axial-vector meson, we consider $D'_{s1}$ as an illustration which receives contributions from $D^{*0}K^+,D^{*+}K^0,D_s^{*+}\eta$  and $D_0^{*0}K^+,D_0^{* +}K^0,D^{*+}_{s0}\eta$ intermediate states (see Fig. \ref{fig:Scalar}(b)).
The full $D'_{s1}$ propagator reads
\be
{-i(g_{\mu\nu}-v_\mu v_\nu)\over 2(v\cdot k-\Delta_{S}-{1\over 4}\Delta M_S-\tilde\Delta_s) -[2\,\Pi_{D^*K}(\omega_{D^*})+{2\over 3}\,\Pi_{D^*_s\eta}(\omega_{D_s^*})+2\,\Pi'_{D_0^*K}(\omega'_{D_0^*})+{2\over 3}\,\Pi'_{D_{s0}^*\eta}(\omega'_{D_{s0}^*})]}, \non \\
\en
where $\omega_{D^*}=v\cdot k+M_{D'_{s1}}-M_{D^*}-{1\over 4}\Delta M_D$ and $\omega'_{D_0^*}=v\cdot k+M_{D'_{s1}}-M_{D^*_0}-(M_{D_0^*}-M_D)+{3\over 4}\Delta M_D$.

The loop integrals in Eqs. (\ref{eq:loop})  and (\ref{eq:loop2}) have the expressions \cite{Falk,Boyd,Stewart}
\be \label{eq:Pi}
\Pi(\omega) &=& \left({2h^2\over f_\pi^2}\right){\omega\over 32\pi^2}\left[ (m^2-2\omega^2){\rm ln}{\Lambda^2\over m^2}-2\omega^2+4\omega^2 F\left(-{m\over \omega}\right)\right]
\en
and \cite{Mehen,Alhakami}
\be \label{eq:Pi'}
\Pi'(\omega) &=& \left({2g'^2\over f_\pi^2}\right){\omega\over 32\pi^2}\left[ (3m^2-2\omega^2){\rm ln}{\Lambda^2\over m^2}-{10\over 3}\omega^2+4m^2+4(\omega^2-m^2) F\left(-{m\over \omega}\right)\right],
\en
respectively, with
\be \label{eq:F}
F\left({1\over x}\right)=\cases{ {\sqrt{x^2-1}\over x}\,{\rm ln}(x+\sqrt{x^2-1}), & $|x|\geq 1$ \cr
-{\sqrt{1-x^2}\over x}\left[{\pi\over 2}-{\rm tan}^{-1}\left({x\over\sqrt{1-x^2}}\right)\right], & $|x|\leq 1$\,. }
\en
Note that the function $F(-m/\omega)$  can be recast to the form
\be \label{eq:FandG}
F\left(-{m\over \omega}\right)={1\over\omega}\, G(\omega, m)
\en
with
\be
G(\omega,m)=\cases{ \sqrt{\omega^2-m^2}\,[{\rm cosh}^{-1}({\omega\over m})-i\pi], & $\omega>m$ \cr
\sqrt{m^2-\omega^2}\,{\rm cos}^{-1}(-{\omega\over m}), & $\omega^2<m^2$ \cr
-\sqrt{\omega^2-m^2}\,{\rm cosh}^{-1}(-{\omega\over m}), & $\omega < -m$\,. }
\en

The parameter $\Lambda$ appearing in Eqs. (\ref{eq:Pi}) and (\ref{eq:Pi'}) is an arbitrary renormalization scale. In the dimensional regularization approach, the common factor
${2\over \epsilon}-\gamma_E+{\rm ln}4\pi+1$
with $\epsilon=4-n$ can be lumped into the logarithmic term ${\rm ln}(\Lambda^2/m^2)$.  In the conventional practice, it is often to choose $\Lambda\sim \Lambda_\chi$, the chiral symmetry breaking scale of order 1 GeV, to get numerical estimates of chiral loop effects. However, as pointed out in \cite{Cheng:2014}, contrary to the common wisdom, the renormalization scale has to be larger than the chiral symmetry breaking scale of order 1 GeV in order to satisfy the on-shell conditions.
In general, there exist two solutions for $v\cdot\tilde k$ due to two intercepts of the curve with the $v\cdot k$ axis. We shall consider the smaller solution for $v\cdot\tilde k$ as the other solution will yield too large masses. It could be that higher-order heavy quark expansion needs to be taken into account to justify the use of $\Lambda\sim \Lambda_\chi$.

\begin{table}[t]
\caption{Mass shifts ($\delta M\equiv M-{\cal M}$) of heavy scalar mesons calculated in HMChPT. The renormalization scale is taken to be $\Lambda=1.3$ (1.2) GeV for scalar $D$ ($B$) mesons. Bare masses are taken from \cite{Godfrey}. All masses and widths are given in MeV and only the central values are listed here.} \label{tab:predictionI}
\begin{tabular}{|c  c  c c  |  } \hline
~~~Meson~~~  &   ~~~Bare mass ${\cal M}$~~~  & ~~~$\delta M$~~~ & ~~~~~$M$~~~~~~  \\ \hline
$D_{s0}^*$ & 2480  & $-162$ &  2318    \\
$D_0^*$ &  2400  & $-79$ &  2321    \\
$B_{s0}^*$ &   5831  & $-136$ &  5694    \\
$B_{0}^*$ &  5756  &  $-55$ & 5701   \\
\hline
$D_{s1}^{\prime}$ & 2550 & $-98$ & 2452 \\
$D_{1}^{\prime}$ & 2460 & $-41$  & 2419   \\
$B_{s1}^{\prime}$ & 5857 & $-85$ & 5772  \\
$B_{1}^{\prime}$ & 5777 & $-34$ & 5743  \\
\hline
\end{tabular}
\end{table}

In our previous study \cite{Cheng:2014} we argued that
near mass degeneracy and the physical masses of $D_{s0}^*$ and $D_0^*$ cannot be accounted for simultaneously in the approach of HMChPT. In this work we show that near mass degeneracy can be implemented by taking into account the additional contributions from axial-vector heavy mesons to the self-energy diagram Fig. \ref{fig:Scalar}(a) of scalar mesons. Since $\omega_P\sim {\cal M}_{P_0^*}-M_P+\cdots$, $\omega'_{P'_1}\sim {\cal M}_{P_0^*}-M_{P'_1}-(M_{P'_1}-M_P)+\cdots$ and ${\cal M}_{P_0^*}>M_P$, $ {\cal M}_{P_0^*}<M_{P'_1}$, we find numerically that $\Pi'(\omega')$ contributes destructively to the mass shifts.
Moreover, the self-energy of $D_{0}^*$ ($D_{s0}^*$) is sensitive (insensitive) to the coupling $g'$.
Therefore, we can adjust the couplings $h$, $g'$ and the renormalization scale $\mu$ to get a mass degeneracy of $D_{s0}^*$ and $D_0^*$. Take the quark model  of \cite{Godfrey} as an illustration which predicts the bare masses of 2480 and 2400 MeV for $D_{s0}^*$ and $D_0^*$, respectively. We find that $h=0.51$, $g'=0.25$ and $\Lambda=1.27$ GeV will lead to $M_{D_{s0}^*}=2222$ MeV and $M_{D_{0}^*}=2225$ MeV.
By
fixing $h$ and $\Lambda$ and varying the coupling $g'$, we obtain $(M_{D_{s0}^*},M_{D_0^*})=(2338,2241)$, $(2277,2235)$ and $(2222,2225)$ MeV for $g'=0$, 0.15 and 0.25, respectively. This explains why we cannot obtain near degeneracy of $D_{s0}^*$ and $D_0^*$ in our previous work with $g'=0$. Hence, although the coupling $g'$ which characterizes the interaction of $P^*_0P'_1$ with the Goldstone boson is small, it does play a crucial role for achieving near degeneracy.
The absolute mass is not an issue as we can scale up the unknown mass $M_0$ by an amount of 96 MeV for scalar and axial-vector heavy mesons, i.e. $P_{0a}^*$ and $P'_{1a}$ in Eq. (\ref{eq:massrel}), so that we have
$M_{D_{s0}^*}=2318$ MeV and $M_{D_{0}^*}=2321$ MeV.

Using the same set of values for the parameters $h$ and $g'$ but  $\Lambda=1.18$ GeV for even-parity $B$ mesons, we see from Table \ref{tab:predictionI} that the closeness of $B_{s0}^*$ and $B_0^*$ masses also holds in the scalar $B$ sector.
The predicted masses for axial-vector charmed mesons $M_{D'_{s1}}=2452$ MeV and $M_{D'_{1}}=2419$ MeV are in agreement with the respective measured masses (see Table \ref{tab:pwavecharm}): $2459.5\pm0.6$ MeV and $2427\pm36$ MeV.

The coupling constant $h$ at tree level can be extracted from the measured widths of $D_0^{*0},D_0^{*\pm}$ and ${D'}_1^{0}$. We find the values $h=0.60\pm0.07$ from $D_0^*(2400)^0$, $0.514\pm0.017$ from $D_0^*(2400)^\pm$ and $0.79\pm0.17$ from $D'_1(2430)^0$. It is obvious that $D_0^*(2400)^\pm$ gives the best determination of $h$ as its mass and width measured recently by LHCb \cite{LHCb:D0pm} are more accurate than $D_0^{*0}$ and ${D'}_1^{0}$. Our result of $h=0.508$ agrees well with the LHCb experiment. As for the coupling $g'$, its magnitude and even the sign relative to the coupling $g$ in Eq. (\ref{eq:Lagrangian}) are unknown. A recent lattice calculation yields  $h=0.84(3)(2)$ and $g'=-0.122(8)(6)$ \cite{Blossier}.

Several remarks are in order.
(i) The chiral loop calculations presented here are meant to demonstrate that the mass difference between strange heavy scalar mesons and their non-strange partners can be substantially reduced by self-energy contributions. Near mass degeneracy in the scalar charm sector will imply the same phenomenon in the scalar $B$ system.
(ii) For the bare masses of even-parity heavy mesons, we have used the relativistic quark model of \cite{Godfrey} as an illustration. If we use the quark model predictions from \cite{Di Pierro}, near degeneracy can be implemented by employing $h=0.52$ and $g'=0.28$ and similar $\Lambda$ given before. (iii) So far we have focused on the closeness of $D_{s0}^*$ and $D_0^{*0}$ masses . The measured mass of $D_0^{*\pm}$ by LHCb is greater than $D_0^{*0}$ and $D_{s0}^*$ by order 30 MeV. More precise measurement of the $D_0^{*0}$ mass is certainly needed. If it turns out that the mass of $D_0^{*0}$ is larger than that of $D_{s0}^*$, then we need a larger $|g'|$ to render $D_0^{*0}$ heavier than $D_{s0}^*$. (iv) There exist other contributions which are not considered in  Fig. \ref{fig:Scalar}, for example,  the intermediate states $D^{0}\ov K^{*0}$, $D_s\eta',\cdots$ for the self-energy diagram of $D_{s0}^*$.
(v)  The calculated masses are sensitive to
the choice of the renormalization scale $\Lambda$.  Of course, physics should be independent of the renormalization scale. However, this issue cannot be properly addressed in the phenomenological model discussed here. In view of this, we should not rely on the chiral loop calculations to make quantitative predictions on the scalar meson masses. For this purpose, we will turn to heavy quark symmetry in Sec. IV below.

\section{Comparison with other works in HMChPT}

Self-energy hadronic loop corrections to $0^+$ and $1^{\prime+}$ heavy mesons have been discussed in the literature within the framework of HMChPT \cite{Becirevic,Fajfer,Brdnik,Mehen,Alhakami,Ananth,Guo,Cheng:2014}.
Guo, Krewald and Mei\ss ner (GKM) \cite{Guo}  considered three different models for calculations. Models I and III correspond to non-derivative and derivative couplings of the scalar meson with two pseudoscalar mesons, while Model II is based on HMChPT. However, our HMChPT results are different from GKM. This can be traced back to the loop integral $\Pi(v\cdot k)$ which has the following expression in \cite{Guo}
\be \label{eq:otherPi}
\Pi(\omega)
&=& =\left({2h^2\over f_\pi^2}\right){m^2\over 32\pi^2}\left[-2\omega\,{\rm ln}{\Lambda^2\over m^2}-2\omega+4 G(\omega,m)\right]  \non \\
&=& \left({2h^2\over f_\pi^2}\right){m^2\over 32\pi^2}\left[-2\omega\,{\rm ln}{\Lambda^2\over m^2}-2\omega+4\omega F\left(-{m\over \omega}\right)\right],
\en
where use of Eq. (\ref{eq:FandG}) has been made.
Comparing with Eq. (\ref{eq:Pi}), we see that the $m^2$ coefficient in Eq. (\ref{eq:otherPi}) should be replaced by $\omega^2$ and a term $(m^2/\omega){\rm ln}(\Lambda^2/m^2)$ in $[\cdots]$ is missing.  Contrary to the claim made by GKM, the self-energy contribution should not vanish in the chiral limit. Some other detailed comparisons are referred to \cite{Cheng:2014}. Contributions from axial-vector heavy mesons to the self-energy of scalar mesons were also not considered in this work.

\begin{table}[t]
\caption{Same as Table \ref{tab:predictionI} except that the propagator $i/(2v\cdot k)$ is used for all heavy meson states and $v\cdot k$ is set to be zero inside the loop integral calculation.} \label{tab:predictionII}
\begin{tabular}{|c  c  c c  |  } \hline
~~~Meson~~~  &   ~~~Bare mass ${\cal M}$~~~  & ~~~$\delta M$~~~ & ~~~~~$M$~~~~~  \\ \hline
$D_{s0}^*$ & 2480  & $-117$ &  2363    \\
$D_0^*$ &  2400  & $-44$ &  2356   \\
$B_{s0}^*$ &   5831  & $-63$ &  5768    \\
$B_{0}^*$ &  5756  &  $-27$ & 5729    \\
\hline
$D_{s1}^{\prime}$ & 2550 & $-48$ & 2502  \\
$D_{1}^{\prime}$ & 2460 & $-22$  & 2438   \\
$B_{s1}^{\prime}$ & 5857 & $-40$ & 5817   \\
$B_{1}^{\prime}$ & 5777 & $-9$ & 5768  \\
\hline
\end{tabular}
\end{table}

Mehen and Springer \cite{Mehen} have systematically studied chiral loop corrections to the masses of scalar and axial-vector heavy mesons. For the propagators of heavy mesons, they only keep the $i/(2v\cdot k)$ terms and neglect all mass splittings such as $\Delta_S$, $\Delta M_{P,S}$ and residual masses such as $\Delta_a$ and $\tilde\Delta_a$ in Eqs. (\ref{eq:Pprop}) and (\ref{eq:Sprop}).  Then they set  $v\cdot k=0$ inside the loop integral. As a consequence, $\omega=M_{\rm ext}-M_{\rm int}$, the mass difference between external and internal heavy meson states.
If we follow this prescription and repeat the calculations presented in the last section, we will obtain different results. The physical masses of scalar charmed mesons read
\be \label{eq:physmass1}
M_{D_{s0}^*}={\cal M}_{D_{s0}^*}+(v\cdot k)_{D_{s0}^*},  \qquad M_{D_{0}^*}={\cal M}_{D_{0}^*}+(v\cdot k)_{D_{0}^*},
\en
with (see Eq. (\ref{eq:FforDs0}))
\be
 (v\cdot k)_{D_{s0}^*} &=& {1\over 2}{\rm Re}\left[2\,\Pi_{DK}(\omega_D)+{2\over 3}\,\Pi_{D_s\eta}(\omega_{D_s} )
+2\,\Pi'_{D'_1K}(\omega_{D'_1})+{2\over 3}\,\Pi'_{D'_{s1}\eta}(\omega_{D'_{s1}})\right], \non \\
 (v\cdot k)_{D_{0}^*} &=& {1\over 2}{\rm Re}\Bigg[{3\over 2}\,\Pi_{D\pi}(\omega_D)+{1\over 6}\,\Pi_{D\eta}(\omega_{D})+\Pi_{D_sK}(\omega_{D_s}) \non \\
&& +{2\over 3}\,\Pi'_{D'_1\pi}(\omega_{D'_1})+{1\over 6}\,\Pi'_{D'_{1}\eta}(\omega_{D'_{1}})+\Pi'_{D'_{s1}K}(\omega_{D'_{s1}})\Bigg],
\en
and $\omega_D={\cal M}_{D_{(s)0}^*}-M_D$, $\omega_{D'_1}={\cal M}_{D_{(s)0}^*}-M_{D'_1}$, $\cdots$ etc. The results of calculations are summarized in Table \ref{tab:predictionII}. It is clear that while near degeneracy is achieved for scalar charmed mesons, it is not the case for scalar $B$ mesons.

Note that in the work of \cite{Mehen} and \cite{Alhakami}, the bare mass ${\cal M}$ in the mass relation $M={\cal M}+\Pi$ is expressed in terms of the unknown low-energy constants in the effective chiral Lagrangian which are fitted to the measured masses of odd- and even-parity charmed mesons. In our work, we have $M={\cal M}+{1\over 2}{\rm Re}\Pi$ (see Eq. (\ref{eq:massPi})) with ${\cal M}$ being inferred from the quark model. We use full propagators to evaluate the mass shifts.
Based on the HMChPT framework of Mehen and Springer \cite{Mehen},  Alhakami \cite{Alhakami} has shown that the nonstrange and strange scalar $B$ mesons are nearly degenerate and the mass difference between $B_0^*$ and $B_{s0}^*$ is $\sim$ 8 MeV. However, the renormalization scale is chosen to be $\Lambda=317$ MeV in \cite{Alhakami} which is too small according to the spirit of ChPT.

\section{Masses of $B^*_{(s)0}$ and $B'_{(s)1}$ mesons}
The states $B^*_{(s)0}$ and $B'_{(s)1}$ are yet to be observed.
The predictions of their masses in the literature are collected in Table \ref{tab:literature}.
They can be classified into the following categories:
(i) relativistic quark potential model  \cite{Di Pierro,Godfrey,Swanson,Liu,Ebert,Zhu,Godfrey:2016},
(ii) nonrelativistic quark model \cite{Lu,Ortega},
(iii) heavy meson chiral perturbation theory \cite{Bardeen,Colangelo,Cheng:2014},
(iv) unitarized chiral perturbation theory \cite{Rincon,Cleven:2010aw,Guo:dynamic,Guo:Ds1,Altenbuchinger,Cleven:2014},
(v) lattice QCD \cite{Gregory,Koponen,Lang:Bs},
(vi) potential model with one loop corrections \cite{Lee,Swanson,Albaladejo},
(vii) QCD sum rules \cite{ZGWang},
and (viii) others, such as the nonlinear chiral SU(3) model \cite{Lutz:2003fm}, the semi-relativistic quark potential model \cite{Matsuki}, the MIT bag model \cite{Orsland},  the mixture of conventional $P$-wave quark-antiquark states with four-quark components \cite{Vijande}, the chiral quark-pion Lagrangian with strong coupled channels \cite{Badalian}, heavy quark symmetry and the assumption of flavor independence of mass differences between $0^+$ and $0^-$ states in \cite{Dmitrasinovic:2012zz}.

It is clear from Table \ref{tab:literature} that  the mass difference between strange and non-strange scalar $B$ mesons predicted by the relativistic quark models is of order $60-110$ MeV,  and that $B_{s0}^*$ is above the $BK$ threshold.
As shown in Sect. II, the closeness of $B_0^*$ and $B_{s0}^*$ masses is expected in view of the near degeneracy observed in the charm sector.

\begin{table}[!]
\centering \caption{Predicted masses (in MeV) of $B^*_{(s)0}$ and $B'_{(s)1}$ mesons in the literature. } \label{tab:literature}
\begin{tabular}{|  l c c c c c | }\hline
 \hskip 1.0 cm &  & $B_0^*$ & $B'_1$ & $B_{s0}^*$ & $B'_{s1}$   \\ \hline
 Relativistic quark model & & & & & \\
 Di Pierro {\it et al.} & \cite{Di Pierro} & 5706 & 5742 & 5804 & 5842   \\
 Godfrey {\it et al.} (2016) & \cite{Godfrey:2016} & 5720 & 5738 & 5805 & 5822   \\
 Lakhina {\it et al.} & \cite{Swanson} & 5730 & 5752 & 5776 & 5803   \\
 Liu {\it et al.} & \cite{Liu} & 5749 & 5782 & 5815 & 5843   \\
 Ebert {\it et al.} & \cite{Ebert} & 5749 & 5774 & 5833 & 5865   \\
 Sun {\it et al.} & \cite{Zhu} & 5756 & 5779 & 5830 & 5858   \\
 Godfrey {\it et al.} (1991) & \cite{Godfrey} & 5756 & 5777 & 5831 & 5857   \\ \hline
 Nonrelativistic quark model & & & & & \\
 Lu {\it et al.} & \cite{Lu} & 5683 & 5729 & 5756 & 5801   \\
 Ortega {\it et al.} & \cite{Ortega} & & & 5741 & 5858   \\ \hline
 HMChPT & & & & & \\
 Bardeen {\it et al.} & \cite{Bardeen} & $5627\pm35$ & $5674\pm35$ & $5718\pm35$ & $5765\pm35$    \\
 Colangelo {\it et al.} & \cite{Colangelo} & $5708.2\pm22.5$ \hskip 2 cm & $5753.3\pm31.1$ \hskip 2 cm & ~$5706.6\pm1.2$~ \hskip 2 cm  & $5765.6\pm1.2$ \hskip 2 cm   \\
 Cheng {\it et al.} & \cite{Cheng:2014} & $5715\pm22+\delta\Delta_S$  & $5752\pm31+\delta\Delta_S$ &  $5715\pm1+\delta\Delta_S$ & $5763\pm1+\delta\Delta_S$ \hskip 2 cm   \\ \hline
 Unitarized ChPT & & & & & \\
 Torres-Rincon {\it et al.} & \cite{Rincon}~~& 5530 & 5579 & 5748 & 5799  \\
 Guo {\it et al.} & \cite{Guo:dynamic,Guo:Ds1} & Two poles & & $5725\pm39$ & $5778\pm7$   \\
 Cleven {\it et al.} & \cite{Cleven:2014} & & & $5625\pm45$ & $5671\pm45$   \\
 Altenbuchinger {\it et al.} & \cite{Altenbuchinger} & & & $5726\pm28$ & $5778\pm26$   \\
 Cleven {\it et al.} & \cite{Cleven:2010aw} & & & $5696\pm40$ & $5742\pm40$    \\ \hline
 Lattice  QCD & & & & & \\
 Gregory {\it et al.} (HPQCD) & \cite{Gregory} & & & $5752\pm30$ & $5806\pm30$   \\
 Koponen {\it et al.} (UKQCD) & \cite{Koponen} & & & $5760\pm9$ & $5807\pm9$   \\
 Lang {\it et al.} & \cite{Lang:Bs} & & & $5713\pm22$ & $5750\pm26$   \\
 \hline
 Chiral loop corrections & & & & & \\
 I. W. Lee {\it et al.} & \cite{Lee} & 5637 & 5673 & 5634 & 5672   \\
 Albaladejo {\it et al.} & \cite{Albaladejo} & & & $5709\pm8$ & $5755\pm8$   \\ \hline
 QCD sum rules & & & & & \\
 Z. G. Wang &  \cite{ZGWang} & $5720\pm50$ & $5740\pm50$ & $5700\pm60$ & $5760\pm60$  \\ \hline
 Others & & & & & \\
 Kolomeitsev {\it et al.} & \cite{Lutz:2003fm} & 5526 &  5590 & 5643 & 5690   \\
 Matsuki {\it et al.} & \cite{Matsuki} & 5592 & 5649 & 5617 & 5682  \\
 Orsland {\it et al.} & \cite{Orsland} & 5592 & 5671 & 5667 & 5737   \\
 Vijande {\it et al.} & \cite{Vijande} & 5615 & & 5679 & 5713   \\
 Badalian {\it et al.} & \cite{Badalian} & $5675\pm20$ & $5725\pm20$ & $5710\pm15$ & $5730\pm15$   \\
 L\"ahde {\it et al.} & \cite{Riska} & 5678 & 5686 & 5781 & 5795   \\
 Dmitra\v{s}inovi\'{c} & \cite{Dmitrasinovic:2012zz} & $5718\pm25$ & $5732\pm25$ & $5719\pm25$ & $5765\pm25$   \\
 \hline
\end{tabular}
\end{table}


Since this section based on heavy quark symmetry was already discussed in details in our previous work \cite{Cheng:2014}, we shall recapitulate the main points and update the numerical results as the masses of charged and neutral $D_0^*$ under the current measurement are somewhat different.

In the heavy quark limit, the two parameters $\Delta_S$ and $\lambda_2^S$ defined in Eqs. (\ref{eq:DeltaS}) and (\ref{eq:lambda2}), respectively,
are independent of the heavy quark flavor,
where $\Delta_S$ measures the spin-averaged mass splitting between the scalar doublet $(P'_1,P^*_0)$ and the pseudoscalar doublet $(P^*,P)$ and $\lambda_S$ is the mass splitting between spin partners, namely $P'_1$ and $P^*_0$, of the scalar doublet. From the data listed in Table \ref{tab:pwavecharm}, we are led to
\be \label{eq:DeltaScu}
\Delta_S^{(c\bar u)}=426\pm34\,{\rm MeV},\qquad \lambda_{2}^{S(c\bar u)}=(360\pm95\,{\rm MeV})^2
\en
for the scalar $c\bar u$ meson,
\be \label{eq:DeltaScd}
\Delta_S^{(c\bar d)}=434\pm29\,{\rm MeV},\qquad \lambda_{2}^{S(c\bar d)}=(302\pm75\,{\rm MeV})^2
\en
for the scalar $c\bar d$ meson and
\be  \label{eq:DeltaScs}
\Delta_S^{(c\bar s)}=347.7\pm0.6\,{\rm MeV},\qquad \lambda_{2}^{S(c\bar s)}=(411.5\pm1.7\,{\rm MeV})^2
\en
for the $c\bar s$ meson, where we have assumed that the charged $D^{\prime\pm}_1$ has the same mass as $D^{\prime 0}_1$.
The large errors associated with $\Delta_S$ and $\lambda_2^S$ in the non-strange scalar meson sector reflect the experimental difficulty in identifying the broad states. The light quark flavor dependence of $\Delta_S$ and $\lambda_2^S$ shown in Eqs. (\ref{eq:DeltaScu})$-$(\ref{eq:DeltaScs})
indicates SU(3) breaking effects.

Note that the parameter $\Delta_S$ can be decomposed in terms of the mass differences between $0^+$ and $0^-$ states and between $1'^+$ and $1^-$ states defined by
\be
&& \Delta_{0}^{(c\bar s)}\equiv M_{D_{s0}^*}-M_{D_s}, \qquad \Delta_{0}^{(c\bar q)}\equiv M_{D_{0}^*}-M_{D}, \non \\
&& \Delta_{1}^{(c\bar s)}\equiv M_{D'_{s1}}-M_{D_s^*}, \qquad \Delta_{1}^{(c\bar q)}\equiv M_{D'_{1}}-M_{D^*}.
\en
We have
\be
\Delta_S^{(c\bar s)}={1\over 4}\left(3\Delta_{1}^{(c\bar s)}+\Delta_{0}^{(c\bar s)}\right), \qquad
\Delta_S^{(c\bar q)}={1\over 4}\left(3\Delta_1^{(c\bar q)}+\Delta_0^{(c\bar q)}\right).
\en
If we follow \cite{Dmitrasinovic:2012zz} to assume the heavy flavor independence of $\Delta_0$ and $\Delta_1$, namely,
\be
\Delta_{0}^{(b\bar s)}=\Delta_{0}^{(c\bar s)}, \qquad \Delta_{0}^{(b\bar q)}=\Delta_{0}^{(c\bar q)}, \qquad \Delta_{1}^{(b\bar s)}=\Delta_{1}^{(c\bar s)}, \qquad \Delta_{1}^{(b\bar q)}=\Delta_{1}^{(c\bar q)},
\en
we will obtain (in MeV)
\be
&& M_{(B_0^*)^0}=5733\pm29, \qquad ~M_{(B'_1)^0}=5745\pm36, \non \\ && M_{(B_0^*)^\pm}=5765\pm6,  \qquad ~~M_{(B'_1)^\pm}=5741\pm36,  \\
&&M_{B_{s0}^*}=5716.2\pm0.6, \qquad M_{B'_{s1}}=5762.7\pm1.8 \,.  \non
\en
Our results for $B^{*0}_{0}$ and $B^{\prime 0}_{1}$ masses are slightly different from those shown in the last row of Table \ref{tab:literature}.  We see that the central values of $B^{*0}_{0}$ and $B^*_{s0}$ are not very close. It appears that the assumption of heavy quark flavor independence of $\Delta_0$ and $\Delta_1$ is probably too strong. Hence, we shall follow \cite{Colangelo} to assume the heavy quark flavor independence of $\Delta_S$ and $\lambda_2^S$,
\be \label{eq:HQSrel}
\Delta_S^{(b\bar q)}=\Delta_S^{(c\bar q)},\quad \Delta_S^{(b\bar s)}=\Delta_S^{(c\bar s)} \quad
\lambda_{2}^{S(b\bar q)}=\lambda_{2}^{S(c\bar q)}, \quad \lambda_{2}^{S(b\bar s)}=\lambda_{2}^{S(c\bar s)}.
\en
In this case,
we find the results (in MeV)
\be \label{eq:BmassHQS}
&& M_{(B_0^*)^0}=5705\pm52, \qquad ~M_{(B'_1)^0}=5751\pm40, \non \\
&& M_{(B_0^*)^\pm}=5724\pm41, \qquad M_{(B'_1)^\pm}=5755\pm34,  \\
&&M_{B_{s0}^*}=5707\pm1, \qquad\quad ~M_{B'_{s1}}=5766\pm1 \,.  \non
\en
Predictions along this approach were first obtained in \cite{Colangelo}.
Evidently, mass degeneracy in the scalar $D$ sector will repeat itself in the $B$ system through heavy quark symmetry. A scrutiny of Table \ref{tab:literature} shows that this degeneracy is respected only in few of the predictions such as \cite{Cheng:2014}, \cite{Alhakami}, \cite{Colangelo}, \cite{Lee},  and \cite{Dmitrasinovic:2012zz}.

Though the relations in Eq. (\ref{eq:HQSrel}) are valid in the heavy quark limit, they do receive $1/m_Q$ and QCD corrections.  The leading QCD correction to the relation $\lambda_2^{S(b\bar q)}=\lambda_2^{S(c\bar q)}$ is given by \cite{Amoros:1997rx}
\be \label{eq:lambda2corr}
\lambda_{2}^{S(b\bar q)}=\lambda_{2}^{S(c\bar q)}\left( {\alpha_s(m_b)\over \alpha_s(m_c)}\right)^{9/25},
\en
and a similar relation with the replacement of $\bar q$ by $\bar s$.
We then obtain
\be
\lambda_2^{S(b\bar u)}=(326\pm86\,{\rm MeV})^2, \quad \lambda_2^{S(b\bar d)}=(273\pm68\,{\rm MeV})^2, \quad \lambda_2^{S(b\bar s)}=(372.6\pm1.5\,{\rm MeV})^2,
\en
and the masses (in MeV)
\be \label{eq:BmassQCD}
&& M_{(B^*_{0})^{0}}=5711\pm49, \quad~ M_{(B'_{1})^{0}}= 5748\pm39, \non \\
&& M_{(B^*_{0})^{\pm}}=5728\pm38, \quad M_{(B'_{1})^{\pm}}= 5754\pm32, \non \\
&& M_{B_{s0}^*}= 5715\pm1, \qquad ~M_{B'_{s1}}= 5763\pm1.
\en
Comparing with Eq. (\ref{eq:BmassHQS}) we see that QCD corrections will mainly enhance the masses of $B_0^*$ and $B_{s0}^*$ by an amount of a few MeV.

We next proceed to the corrections to $\Delta_S^{(b)}=\Delta_S^{(c)}$. In heavy quark effective theory, it follows that (see \cite{Cheng:2014} for details)
\be \label{eq:1/mQcorr}
\Delta_S^{(b)}=\Delta_S^{(c)}+\delta\Delta_S\equiv \Delta_S^{(c)}+(\lambda_1^S-\lambda_1^H)\left( {1\over 2m_c}-{1\over 2m_b}\right).
\en
While the parameter $\lambda_1^H$ is calculable,
the other parameter $\lambda_1^S$ is unknown. In \cite{Cheng:2014}
the $1/m_Q$ correction is estimated to be
\be
\delta\Delta_S\sim {\cal O}(-35)\,{\rm MeV},
\en
to be compared with the estimate of $-50\pm25$ MeV in \cite{Mehen}.

 As pointed out in \cite{Cheng:2014}, so long as $\delta\Delta_S$ is not large in magnitude, the $1/m_Q$ correction to the relation $\Delta _S^{(b)}=\Delta _S^{(c)}$ [see Eq. (\ref{eq:1/mQcorr})], to a very good approximation, amounts to lowering the masses of $B^*_{(s)0}$ and $B'_{(s)1}$ by an equal amount of $|\delta\Delta_S|$:
\be \label{eq:Bmass1/mQ}
&& M_{(B^*_{0})^{0}}=5711\pm49\,{\rm MeV}+\delta\Delta_S, \quad~ M_{(B'_{1})^{0}}= 5748\pm39\,{\rm MeV}+\delta\Delta_S, \non \\
&& M_{(B^*_{0})^{\pm}}=5728\pm38\,{\rm MeV}+\delta\Delta_S, \quad M_{(B'_{1})^{\pm}}= 5754\pm32\,{\rm MeV}+\delta\Delta_S, \non \\
&& M_{B_{s0}^*}= 5715\pm1\,{\rm MeV}+\delta\Delta_S, \qquad ~M_{B'_{s1}}= 5763\pm1\,{\rm MeV}+\delta\Delta_S.
\en
Evidently, the mass degeneracy between $B_{s0}^*$ and $B_0^*$ shown in Eq. (\ref{eq:BmassHQS}) implied by heavy quark symmetry is not spoiled by $1/m_Q$ and QCD corrections.

Several remarks are in order. (i) There exists an empirical mass relation
\be \label{eq:Dmassdifrel}
M_{D'_{s1}}-M_{D_{s0}^*}\simeq M_{D^*}-M_D\simeq M_{D_s^*}-M_{D_s},\qquad\quad (141.8, 141.4, 143.8)~{\rm MeV}
\en
for charmed mesons.
Therefore, one will expect the mass difference relation
\be \label{eq:Bmassdifrel}
M_{B'_{s1}}-M_{B_{s0}^*}\simeq M_{B^*}-M_B\simeq M_{B_s^*}-M_{B_s}
\en
also valid in the $B$ sector. The mass difference between $B'_{s1}$ and $B_{s0}^*$ in Eq. (\ref{eq:Bmass1/mQ}) is independent of $\delta\Delta_S$; its numerical value $48\pm1$ MeV indeed respects the relation (\ref{eq:Bmassdifrel}). It is interesting that this mass relation is not obeyed by any of the existing relativistic quark models  which predict a smaller mass difference $M_{B'_{s1}}-M_{B^*_{s0}}\sim 20-30$ MeV (see Table \ref{tab:literature}). (ii) In the charm sector we have the hierarchy $M_{D'_{s1}}-M_{D^*_{s0}}>M_{D'_1}-M_{D^*_0}$. It is natural to generalize this to the $B$ sector: $M_{B'_{s1}}-M_{B^*_{s0}}>M_{B'_1}-M_{B^*_0}$.
Indeed, the mass splittings of the $j={1\over 2}$ doublet in strange and nonstrange sectors satisfy the relation (see \cite{Cheng:2014})
\be
{M_{B'_{s1}}-M_{B_{s0}^*} \over M_{B'_{1}}-M_{B_{0}^*} }={\lambda_2^{S(b\bar s)}\over \lambda_2^{S(b\bar q)} }, \qquad {M_{D'_{s1}}-M_{D_{s0}^*} \over M_{D'_{1}}-M_{D_{0}^*} }={\lambda_2^{S(c\bar s)}\over \lambda_2^{S(c\bar q)} }.
\en
The relation $M_{B'_{s1}}-M_{B^*_{s0}}>M_{B'_1}-M_{B^*_0}$ is not respected by some of the models listed in Table \ref{tab:literature}.
(iii) The difference 15 MeV between the central values of $M_{B'_{s1}}$ and $M_{B'_1}$ inferred from Eq. (\ref{eq:Bmass1/mQ}) is much smaller than the quark model expectation of $60-100$ MeV (cf Table \ref{tab:literature}).
\footnote{The splitting between strange and nonstrange  axial-vector bottom mesons is of order 19 MeV in the HHChPT approach of \cite{Alhakami}.}
Just as the scalar $B$ mesons, the mass splitting is reduced by the self-energy effects due to strong coupled channels. (iv) The current world-averaged masses of neutral and charged $D_0^*(2400)$ mesons are not the same; they differ by an amount of $33\pm29$ MeV and the charged one is better measured. On the experimental side, it is thus important to have a more precise mass measurement of the scalar charmed meson. If it turns out the $D_0^{*0}$ mass is also of order 2350 MeV,  we find from Eq. (\ref{eq:Bmass1/mQ}) that the $B_0^*$ mass is larger than that of $B_{s0}^*$ by an order of 13 MeV.

Since the masses of $B_{s0}^*$ and $B'_{s1}$ are below the $BK$ and $B^*K$ thresholds, respectively, their widths are expected to be very narrow with the isospin-violating strong decays into $B_s\pi^0$ and $B_s^*\pi^0$. Experimentally, it will be even more difficult to identify the non-strange $B_{0}^*$ and $B'_{1}$ mesons owing to their broad widths.

\section{Conclusions}
The $q\bar q$ quark model encounters a great challenge in describing even-parity mesons. Specifically, it has difficulties in understanding the light scalar mesons below 1 GeV, scalar and axial-vector charmed mesons and $1^+$ charmonium-like state $X(3872)$.
Many studies indicate that quark model results are distorted by the interaction of the even-parity meson with strong coupled channels. In this work, we focus on the near mass degeneracy of scalar charmed mesons, $D_{s0}^*$ and $D_0^{*0}$, and its implications to the $B$ sector.

\begin{itemize}

\item
We work in the framework of HMChPT to evaluate the self-energy diagrams for $P_0^*$ and $P'_1$ heavy mesons.
The approximate mass degeneracy can be qualitatively understood as a consequence of self-energy hadronic loop corrections which will push down the mass of the heavy scalar meson in the strange sector more than that in the non-strange one. Quantitatively, the closeness of $D_{s0}^*$ and $D_0^{*0}$ can be implemented by adjusting the strong couplings $h$ and $g'$ and the renormalization scale $\Lambda$. This in turn implies the mass similarity of $B_{s0}^*$ and $B_0^{*0}$. The $P_0^* P'_1$ interaction with the Goldstone boson characterized by the strong coupling $g'$ is crucial for understanding the phenomenon of near degeneracy.

\item
The self-energy loop diagram has been evaluated in the literature by neglecting mass splittings and residual masses in the heavy meson's propagator. If we follow this prescription, we find near degeneracy in the scalar charm sector but not in the corresponding $B$ system.

\item
The masses of $B^*_{(s)0}$ and $B'_{(s)1}$ mesons (i.e. Eq. (\ref{eq:Bmass1/mQ})) can be deduced from the charm spectroscopy and heavy quark symmetry with corrections from QCD and $1/m_Q$ effects,  for example, $M_{B_{s0}^*}= (5715\pm1)\,{\rm MeV}+\delta\Delta_S$, $M_{B'_{s1}}=(5763\pm1)\,{\rm MeV}+\delta\Delta_S$ with $\delta\Delta_S$ being $1/m_Q$ corrections. We found that the mass difference of 48 MeV between $B'_{s1}$ and $B_{s0}^*$ is larger than the result of $20\sim 30$ MeV predicted by the quark models, whereas the difference of 15 MeV between the central values of $M_{B'_{s1}}$ and $M_{B'_1}$ is much smaller than the quark model expectation of $60-100$ MeV.

\item
The current world-averaged masses of neutral and charged $D_0^*(2400)$ mesons are not the same; they differ by an amount of $33\pm29$ MeV and the charged one is more well measured. Experimentally, it is thus important to have a more precise mass measurement of the scalar charmed meson, especially the neutral one. If it turns out the $D_0^{*0}$ mass is of order 2350 MeV, this means $D^*_0$ is heavier than $D_{s0}^*$ even though the latter contains a strange quark. In this case, we find that the $B_0^*$ mass is larger than that of $B_{s0}^*$ by an order of 13 MeV.

\end{itemize}

\vskip 1.2cm \acknowledgments
We wish to thank Ting-Wai Chiu and Keh-Fei Liu for insightful discussions.
This research was supported in part by the Ministry of Science and Technology of R.O.C. under the Grant No. 104-2112-M-001-022 and by the National Science Foundation of China under the Grant No. 11347027.


\end{document}